\documentclass[11pt]{amsart}
\usepackage{amssymb}
\usepackage{amsmath}
\usepackage{amscd}
\usepackage{amsfonts}
\usepackage{mathrsfs}
\usepackage{stmaryrd}
\usepackage{bbm}
\newtheorem{theorem}{Theorem}[section]
\newtheorem{prop}[theorem]{Proposition}
\newtheorem{defn}[theorem]{Definition}
\newtheorem{lemma}[theorem]{Lemma}
\newtheorem{coro}[theorem]{Corollary}
\newtheorem{prop-def}{Proposition-Definition}[section]

\newtheorem{remark}[theorem]{Remark}

\newtheorem{exam}[theorem]{Example}

\begin{document}
\setlength{\oddsidemargin}{0cm} \setlength{\evensidemargin}{0cm}

\title{L-quadri-algebras}

\author{Ligong Liu}

\address{Chern Institute of Mathematics \& LPMC, Nankai
University, Tianjin 300071, P.R.
China}\email{liuligong@mail.nankai.edu.cn}

\author{Xiang Ni}

\address{Chern Institute of Mathematics \& LPMC, Nankai
University, Tianjin 300071, P.R.
China}\email{xiangn$_-$math@yahoo.cn}

\author{Chengming Bai$^*$}

\address{Chern Institute of Mathematics \& LPMC, Nankai University,
Tianjin 300071, P.R. China} \email{baicm@nankai.edu.cn}\thanks{$^*$
Corresponding author}

\def\shorttitle{L-quadri-algebras}

\begin{abstract}
Quadri-algebras introduced by Aguiar and Loday are a class of
remarkable Loday algebras. In this paper, we introduce a notion of
L-quadri-algebra with 4 operations satisfying certain generalized
left-symmetry, as a Lie algebraic analogue of quadri-algebra such
that the commutator of the sum of the 4 operations is a Lie algebra.
Any quadri-algebra is an L-quadri-algebra. Moreover,
L-quadri-algebras fit into the framework of the relationships
between Loday algebras and their Lie algebraic analogues, extending
the well known fact that the commutator of an associative algebra is
a Lie algebra. We also give the close relationships between
L-quadri-algebras and some interesting structures like Rota-Baxter
operators, classical Yang-Baxter equation, some bilinear forms
satisfying certain conditions.

\end{abstract}



\maketitle


\baselineskip=18pt

\section{Introduction}
\setcounter{equation}{0}
\renewcommand{\theequation}
{1.\arabic{equation}}

Quadri-algebras, which were introduced by Aguiar and Loday
(\cite{AL}), are a remarkable class of Loday algebras. The Loday
algebras (\cite{Va}, or ABQR operad algebras in \cite{EG}) refer to
the series of algebras with a common property of ``splitting
associativity" (\cite{Lo3}), which the first and maybe the most
important class of Loday algebras, namely, the  dendriform algebras,
were introduced by Loday (\cite{Lo1}) in 1995 with motivation from
algebraic $K$-theory. The other Loday algebras include dendriform
trialgebras (\cite{E,LR3}), $NS$-algebras (\cite{Le2}),
octo-algebras (\cite{Le3}), ennea-algebras (\cite{Le1}) and
dendriform-Nijenhuis algebras (\cite{Le2}).

In fact, the Loday algebras were introduced due to their own
interesting motivations and then developed as independent algebraic
systems quickly. For example, dendriform algebras have been studied
quite extensively with connections to operads (\cite{Lo3}), homology
(\cite{Fr1}), Hopf algebras (\cite{Cha,LR2}), Lie and Leibniz
algebras (\cite{Fr2}), combinatorics (\cite{LR1}),
arithmetic(\cite{Lo2}) and quantum field theory (\cite{Fo}) and so
on (see \cite{EMP} and the references therein). Quadri-algebras were
introduced due to their close relationships with dendriform algebras
in several ways like the relations between Rota-Baxter operators and
infinitesimal bialgebras (\cite{AL}). They are closely
related to the study of ennea-algebras (\cite{Le1}), operads
(\cite{EG}) and $\mathcal O$-operators (\cite{Bai3}).

In this paper, we introduce a notion of L-quadri-algebra due to
several different motivations.

{\bf (1) Relationships with Loday algebras.} An L-quadri-algebra is
regarded as a Lie algebraic analogue of a quadri-algebra, which is
one of the main motivations to introduce such a notion. It is known
that the commutator of an associative algebra is a Lie algebra.  It
is quite interesting to find that certain commutator of a dendriform
algebra is a pre-Lie algebra (\cite{Cha,Ron}) which has
already been introduced independently in geometry (\cite{Ko,Vi}) and
deformations of associative algebras (\cite{G}) and then appear in
many fields (\cite{Bu} and the references therein). Moreover,
 a close relationship among Lie algebras, associative
algebras, pre-Lie algebras and dendriform algebras is given as
follows in the sense of commutative diagram of categories
(\cite{Cha}):
\begin{equation}\begin{matrix} \mbox{Lie algebra}
&\stackrel{}{\leftarrow} & \mbox{Pre-Lie algebra} \cr \uparrow
&&\uparrow  \cr
 \mbox{Associative algebra} &\stackrel{}{\leftarrow}
& \mbox{Dendriform algebra}\cr
\end{matrix}\end{equation}

Then it is natural to ask whether there are certain similar
structures as of ``Loday algebras" on Lie algebras. In fact, due
to the different motivations, a notion of L-dendriform algebra was
introduced in \cite{BLN} as a Lie algebraic analogue of dendriform
algebra, which extends the commutative diagram (1.1) to the
following commutative diagram:
\begin{equation}\begin{matrix} \mbox{Lie algebra} &\stackrel{}{\leftarrow} &
\mbox{Pre-Lie algebra}& \stackrel{}{\leftarrow} & \mbox{L-dendriform
algebra}\cr & \stackrel{}{\nwarrow} & &\stackrel{}{\nwarrow} &
&\stackrel{}{\nwarrow} &  \cr
 & & \mbox{Associative algebra} &\stackrel{}{\leftarrow}
& \mbox{Dendriform algebra}&\stackrel{}{\leftarrow} &
\mbox{Quadri-algebra}& \cr
\end{matrix}
\end{equation}

As the above commutative diagram (1.2) at the level of associative
algebras extends to the more Loday algebras (like octo-algebras), it
is natural to find the corresponding algebraic structures at the
level of Lie algebras. We will show that the L-quadri-algebras are
chosen in a certain sense such that the following diagram including
both diagrams (1.1) and (1.2) as sub-diagrams is commutative:

{\small \begin{equation}\begin{matrix} \mbox{Lie algebra}
&\stackrel{}{\leftarrow} & \mbox{Pre-Lie algebra}&
\stackrel{}{\leftarrow} & \mbox{L-dendriform algebra}&
\stackrel{}{\leftarrow} &\mbox{L-quadri-algebra} \cr &
\stackrel{}{\nwarrow} & &\stackrel{}{\nwarrow} &
&\stackrel{}{\nwarrow} & & \stackrel{}{\nwarrow}\cr
 & & \mbox{Associative algebra} &\stackrel{}{\leftarrow}
& \mbox{Dendriform algebra}&\stackrel{}{\leftarrow} &
\mbox{Quadri-algebra}&\stackrel{}{\leftarrow} & \mbox{Octo-algebra}
\cr
\end{matrix}
\end{equation}}

{\bf (2) ``Generalized left-symmetry".} A pre-Lie algebra
$(A,\cdot)$ is also called a left-symmetric algebra since the
associator
\begin{equation}
(x,y,z)=x\cdot (y\cdot z)-(x\cdot y)\cdot z,\;\;\forall x,y,z\in A
\end{equation}
is left-symmetric in $x,y$. Note that the associativity is as the
same as the fact that the associator is zero. Furthermore, we show
that for an L-dendriform algebra, there are the ``generalized
associators" with two operations satisfying certain ``generalized
left-symmetry" (\cite{BLN}), whereas a dendriform algebra
satisfies that the generalized associators are zero. In this
sense, the so-called ``common property of splitting associativity"
(\cite{Lo3}) for Loday algebras is understood as the generalized
associators being zero, and it is reasonable to assume that their
Lie algebraic analogues satisfy the generalized left-symmetry.  It
is true for L-quadri-algebras. That is, an L-quadri-algebra is an
algebra with 4 operations satisfying that certain ``generalized
associators" are ``generalized left-symmetric". As a direct
consequence, the commutator of the sum of the 4 operations of an
L-quadri-algebra is a Lie algebra and a quadri-algebra is a
special L-dendriform algebra with the generalized associators
being zero.

{\bf (3) Rota-Baxter operators.} Rota-Baxter operators on
associative algebras first arose in probability
 theory (\cite{Bax}) and later became a subject in combinatorics (\cite{Rot}) and
a lot of fields in mathematics and physics (\cite{E} and the
references therein). A Lie algebraic version of Rota-Baxter
 operator is exactly the operator form of the famous classical
 Yang-Baxter equation (CYBE) (\cite{Se}). As two commuting Rota-Baxter operators on an associative
 algebra or a Rota-Baxter operator on a dendriform algebra give a
 quadri-algebra, which is one of the main motivations to introduce
the notion of quadri-algebra (\cite{AL}, there is a
 similar result for an L-quadri-algebra: three pairwise commuting Rota-Baxter
 operators on a Lie algebra or two commuting Rota-Baxter operators
 on a pre-Lie algebra or a Rota-Baxter operator on an L-dendriform
 algebra give an L-quadri-algebra. The above approach is another
 important motivation to introduce the notion of L-quadri-algebra, as it did for
 quadri-algebras (\cite{AL}).

We briefly introduce the following two motivations involving the Lie
algebras themselves.

{\bf (4) Some nondegenerate  bilinear forms on L-dendriform
algebras.}  L-quadri-algebras are regarded as the underlying
algebraic structures of the  L-dendriform algebras with a
nondegenerate 2-cocycle, just as pre-Lie algebras are the
underlying algebraic structures of the symplectic Lie algebras
(\cite{Chu}) and L-dendriform algebras are the underlying
algebraic structures of the Hessian structures (\cite{Sh}), that
is, the pre-Lie algebras with a nondegenerate 2-cocycle.

{\bf (5) Classical Yang-Baxter equation and $\mathcal
O$-operators.} In fact, pre-Lie algebras can be regarded as the
algebraic structures behind the CYBE which plays an important role
in integrable systems and  quantum groups (\cite{CP} and the
references therein). It could be seen more clearly in terms of
$\mathcal O$-operators of a Lie algebra which was a generalization
of (the operator form of) the CYBE in a Lie algebra
(\cite{Se,Ku}). Moreover, the CYBE and the Rota-Baxter operators
on a Lie algebra are understood as the special $\mathcal
O$-operators which correspond to the co-adjoint representation and
the adjoint representation respectively (\cite{Bai1}). In terms of
$\mathcal O$-operators, there are analogues of the CYBE in pre-Lie
algebras (\cite{Bai2}, $S$-equation) and L-dendriform algebras
(\cite{BLN}, $LD$-equation). We will show that L-quadri-algebras
are the algebraic underpinning of the $LD$-equation and $\mathcal
O$-operators on L-dendriform algebras. In fact, the axioms of an
L-quadri-algebra are equivalent to the fact that the identity map
is an $\mathcal O$-operator (see Theorem 3.17 for details).

The paper is organized as follows. In Section 2, we recall some
basic facts on pre-Lie algebras and L-dendriform algebras. In
Section 3, we introduce the notion of L-quadri-algebra and then
study some basic properties of L-quadri-algebras in terms of the
$\mathcal O$-operators of L-dendriform algebras. Their
relationships with Loday algebras are also given.
 In the
Appendix, in order to be self-contained, we introduce a notion of
L-octo-algebra and then give certain relationships between
L-octo-algebras and L-quadri-algebras.

Throughout this paper, all algebras are finite-dimensional and over
a field of characteristic zero. We also give some notations as
follows. Let $A$ be an algebra with an operation $*$.

(1) Let $L_*(x)$ and $R_*(x)$ denote the left and right
multiplication operator respectively, that is, $L_*(x)y=R_*(y)x=x*y$
for any $x,y\in A$. We also simply denote them by $L(x)$ and $R(x)$
respectively without confusion. Moreover let $L_*, R_*:A\rightarrow
gl(A)$ be two linear maps with $x\rightarrow L_*(x)$ and
$x\rightarrow R_*(x)$ respectively.  In particular, when $A$ is a
Lie algebra, we let ${\rm ad}(x)$ denote the adjoint operator, that
is, ${\rm ad} (x)y=[x,y]$ for any $x,y\in A$.

(2) Let $r=\sum_{i}a_i\otimes b_i\in A\otimes A$. Set
\begin{equation}
r_{12}=\sum_ia_i\otimes b_i\otimes 1, r_{13}=\sum_{i}a_i\otimes
1\otimes b_i,\;r_{23}=\sum_i1\otimes a_i\otimes b_i,
\end{equation}
where $1$ is a symbol which plays a similar role of the unit. The
operation between two $r$s is defined in an obvious way. For
example,
\begin{equation}
r_{12}*r_{13}=\sum_{i,j}a_i*a_j\otimes b_i\otimes b_j,\; r_{13}*
r_{23}=\sum_{i,j}a_i\otimes a_j\otimes
b_i*b_j,\;r_{23}*r_{12}=\sum_{i,j}a_j\otimes a_i* b_j\otimes
b_i.\end{equation}

(3) Let $V$ be a vector space.  Let $\sigma: V\otimes V \rightarrow
V\otimes V$ be the exchanging operator defined as
\begin{equation}\sigma (x \otimes y) = y\otimes x,
\forall x, y\in V.
\end{equation}
 On
the other hand, any $r\in V\otimes V$ can be identified as a linear
map from the dual space $V^*$ to $V$ in the following way:
\begin{equation}
\langle r(v^*) , u^*\rangle = \langle u^*\otimes v^*,r\rangle,
\forall u^*, v^*\in V^*,
\end{equation}
where $\langle,\rangle$ is the ordinary pair between the vector
space $V$ and its dual space $V^*$. Moreover, any invertible
linear map $T:V^*\rightarrow V$ induces a nondegenerate bilinear
form $\mathcal B( , )$ on $V$ by
\begin{equation}
\mathcal B(u,v)=\langle T^{-1}u, v\rangle,\;\;\forall u,v\in V.
\end{equation}

(4) Let $V$ be a vector space. For any linear map $\rho:
A\rightarrow gl(V)$, define a linear map $\rho^* : A \rightarrow
gl(V^*)$ by
\begin{equation}
\langle \rho^*(x)v^*, u\rangle = -\langle v^*, \rho(x)u\rangle,\
\forall x \in A, u\in V, v^*\in V^*.
\end{equation}

\newpage

\section{Pre-Lie algebras and L-dendriform algebras}
\setcounter{equation}{0}
\renewcommand{\theequation}
{2.\arabic{equation}}

\subsection{Pre-Lie algebras and L-dendriform algebras}

\begin{defn} {\rm Let $A$ be a vector space with a
bilinear product $(x,y)\rightarrow x\circ y$. $A$ is called a {\it
pre-Lie algebra} if for any $x,y,z\in A$, the associator
\begin{equation}
(x,y,z)=(x\circ y)\circ z-x\circ (y\circ z) \end{equation} is
symmetric in $x,y$, that is,
\begin{equation}
(x,y,z)=(y,x,z),\;{\rm or}\;{\rm equivalently}\;(x\circ y)\circ
z-x\circ (y\circ z)=(y\circ x)\circ z-y\circ (x\circ z),\;\forall
x,y,z\in A.
\end{equation}}
\end{defn}

\begin{prop} {\rm (\cite{Bu})}\quad Let $(A, \circ)$ be a pre-Lie algebra.

{\rm (1)} The commutator
\begin{equation} [x,y] =
x\circ y - y\circ x,\;\;\forall x,y\in A
\end{equation} defines a Lie algebra $\frak g(A)$, which is called the sub-adjacent Lie algebra of
$A$.

{\rm (2)} $L_\circ$ gives a representation of the Lie algebra $\frak
g(A)$, that is,
\begin{equation}
L_\circ({[x, y]}) = L_\circ(x)L_\circ(y)- L_\circ(y)L_\circ(x),\
\forall x, y\in A.
\end{equation}
\end{prop}

The following conclusion can be obtained directly from Definition 2.1 and Proposition 2.2.

\begin{prop} Let $\frak g $ be a vector space with a bilinear product $\circ$.
Then $(\frak g , \circ)$ is a pre-Lie algebra if and only if $(\frak
g , [,])$ defined by Eq. (2.3) is a Lie algebra and $(L_\circ, \frak
g )$ is a representation of the Lie algebra $(\frak g, [,])$.
\end{prop}

\begin{defn}{\rm (\cite{Lo1,BLN}) Let $A$ be a vector space  with two
bilinear products denoted by $\triangleright$ and $ \triangleleft:
A\otimes A \rightarrow A$.  $(A, \triangleright, \triangleleft)$ is
called an {\it L-dendriform algebra} if for any $x,y,z\in A$,
\begin{equation}
x\triangleright(y\triangleright z)-(x\bullet y)\triangleright z =
y\triangleright(x\triangleright z) - (y\bullet x)\triangleright z,
\end{equation}
\begin{equation}
x\triangleright(y\triangleleft z) -(x\triangleright y)\triangleleft
z= y\triangleleft (x\bullet z)-(y\triangleleft x)\triangleleft z.
\end{equation}
where $x\bullet y=x\triangleright y + x \triangleleft y$. In
particular, if both sides of Eqs. (2.5) and (2.6) are zero, then
$(A,\triangleright, \triangleleft)$ is called a {\it dendriform
algebra}. }\end{defn}

\begin{remark}
{\rm Due to the left-symmetry of $x,y$ in Eq. (2.5), in fact,
there are 3 independent identities for a dendriform algebra, that is,
$$x\triangleright(y\triangleright z)=(x\bullet y)\triangleright z,\;x\triangleright(y\triangleleft z) =(x\triangleright y)\triangleleft
z,\; y\triangleleft (x\bullet z)=(y\triangleleft x)\triangleleft z,\;\;\forall x,y,z\in A.$$
Hence any dendriform algebra is an L-dendriform algebra.}
\end{remark}

\begin{prop}\label{LP} {\rm (\cite{BLN})} Let $(A,\triangleright,
\triangleleft)$ be an L-dendriform algebra.

(1) If we define
\begin{equation}
x\bullet y=x\triangleright y + x \triangleleft y,\forall x, y \in A,
\end{equation}
then $(A,\bullet)$ is a pre-Lie algebra. $(A,\bullet)$ is called the
associated horizontal pre-Lie algebra.

(2) If we define
\begin{equation}
x\circ y=x\triangleright y - y \triangleleft x,\forall x, y \in A,
\end{equation}
then $(A,\circ)$ is a pre-Lie algebra. $(A,\circ)$ is called the
associated vertical pre-Lie algebra.

(3) The associated horizontal and vertical pre-Lie algebras
$(A,\bullet)$ and $(A,\circ)$ of an L-dendriform algebra
$(A,\triangleright, \triangleleft)$ have the same sub-adjacent Lie
algebra $\frak g(A)$ defined by
\begin{equation}
[x,y]=x\triangleright y + x \triangleleft y - y\triangleright x - y
\triangleleft x, \forall x,y\in A.
\end{equation}
This Lie algebra is called the sub-adjacent Lie algebra of the
L-dendriform algebra $(L,\triangleright, \triangleleft)$.
\end{prop}

\subsection{Bimodules and $\mathcal O$-operators}

\begin{defn}{\rm (\cite{Bai2})\quad Let $(A,\circ)$ be a pre-Lie algebra and $V$ be a vector space. Let
$l,r: A\rightarrow gl(V)$ be two linear maps. $(l,r,V)$ is called a
{\it bimodule of $(A,\circ)$} if
\begin{equation}
l(x)l(y)-l(x\circ y)=l(y)l(x)-l(y\circ x),
\end{equation}
\begin{equation}
l(x)r(y)-r(y)l(x)=r(x\circ y)-r(y)r(x),\forall x,y \in A.
\end{equation}}\end{defn}

\begin{prop} {\rm (\cite{BLN})}\quad Let $A$ be a vector space with two
bilinear products denoted by $\triangleright, \triangleleft:
A\otimes A \rightarrow A$.

(1) $(A, \triangleright, \triangleleft)$ is an L-dendriform algebra
if and only if $(A,\bullet)$ defined by Eq. (2.7) is a pre-Lie
algebra and $(L_\triangleright, R_\triangleleft, A)$ is a bimodule.

(2) $(A, \triangleright, \triangleleft)$ is an L-dendriform algebra
if and only if $(A,\circ)$ defined by Eq. (2.8) is a pre-Lie algebra
and $(L_\triangleright, -L_\triangleleft, A)$ is a bimodule.
\end{prop}

\begin{defn}{\rm (\cite{BLN}) Let $(A,\triangleright, \triangleleft)$ be an L-dendriform
algebra and $V$ be a vector space. Let
$l_{\triangleright},r_{\triangleright},l_{\triangleleft},r_{\triangleleft}:A\rightarrow
gl(V)$ be four linear maps.
$(l_{\triangleright},r_{\triangleright},l_{\triangleleft},r_{\triangleleft},
V)$ is called a {\it bimodule of $(A,\triangleright,
\triangleleft)$} if the following five equations hold  (for any $x,
y \in A$)
\begin{equation}
[l_{\triangleright}(x),l_{\triangleright}(y)] =
l_{\triangleright}[x,y];
\end{equation}
\begin{equation}
[l_{\triangleright}(x),l_{\triangleleft}(y)] =
l_{\triangleleft}(x\circ y) +
l_{\triangleleft}(y)l_{\triangleleft}(x);
\end{equation}
\begin{equation}
r_{\triangleright}(x\triangleright y) =
r_{\triangleright}(y)r_{\triangleright}(x) +
r_{\triangleright}(y)r_{\triangleleft}(x) +
[l_{\triangleright}(x),r_{\triangleright}(y)]-
r_{\triangleright}(y)l_{\triangleleft}(x);
\end{equation}
\begin{equation}
r_{\triangleright}(x\triangleleft y) =
r_{\triangleleft}(y)r_{\triangleright}(x) +
l_{\triangleleft}(x)r_{\triangleright}(y) +
[l_{\triangleleft}(x),r_{\triangleleft}(y)];
\end{equation}
\begin{equation}
[l_{\triangleright}(x),r_{\triangleleft}(y)] =
r_{\triangleleft}(x\bullet y) -
r_{\triangleleft}(y)r_{\triangleleft}(x),
\end{equation}
where $x\circ y= x\triangleright y - y\triangleleft x,$ and $
x\bullet y= x\triangleright y + x\triangleleft y$.}\end{defn}

In fact, according to \cite{Sc},
$(l_{\triangleright},r_{\triangleright},l_{\triangleleft},r_{\triangleleft},
V)$ is a bimodule of an L-dendriform algebra $(A, \triangleright,
\triangleleft)$  if and only if the direct sum $A\oplus V$ of the
underlying vector spaces of $A$ and $V$ is turned into an
L-dendriform algebra by defining multiplication in $A\oplus V$ by
(for any $x,y \in A, u,v \in V $)
\begin{equation} (x+u)\triangleright(y+v)=x\triangleright y +
l_{\triangleright}(x)v + r_{\triangleright}(y)u,
\end{equation}
\begin{equation}
(x+u)\triangleleft(y+v)=x\triangleleft y + l_{\triangleleft}(x)v +
r_{\triangleleft}(y)u,
\end{equation}
We denote it by
$A\ltimes_{l_{\triangleright},r_{\triangleright},l_{\triangleleft},r_{\triangleleft}}V$.

The following conclusion involves the notations given by Eq. (1.10).
\begin{prop} {\rm (\cite{BLN})}\quad  Let
$(l_{\triangleright},r_{\triangleright},l_{\triangleleft},r_{\triangleleft},
V)$ be a bimodule of an L-dendriform algebra $(A,\triangleright,\
\triangleleft)$.  Then
$(l_{\triangleright}^*+l_{\triangleleft}^*-r_{\triangleright}^*-r_{\triangleleft}^*,\
r_{\triangleright}^*,r_{\triangleright}^*-l_{\triangleleft}^*,\
-(r_{\triangleright}^*+r_{\triangleleft}^*),V^*)$ is a bimodule of
$(A,\triangleright,\triangleleft)$.
\end{prop}

\begin{defn} {\rm (1) (\cite{Ku}) Let $\rho: \frak g \rightarrow gl(V)$ be a representation of
a Lie algebra $\frak g $. A linear map $T:V \rightarrow \frak g $
is called an {\it $\mathcal {O}$-operator of $\frak g$ associated
to $(\rho, V)$} if $T$ satisfies
\begin{equation}
[T(u), T(v)] = T(\rho(T(u))v - \rho(T(v))u),\forall u,v \in V.
\end{equation}
In particular,  an ${\mathcal O}$-operator $R:\frak g\rightarrow
\frak g$ of a Lie algebra $\frak g$ associated to the adjoint
representation $({\rm ad},\frak g)$ is called a {\it Rota-Baxter
operator of (weight zero) on $\frak g$}, that is,
 $R$ satisfies
\begin{equation}
[R(x),R(y)]=R([R(x),y]+[x,R(y)]),\;\;\forall x,y\in \frak g.
\end{equation}

(2) (\cite{BLN}) Let $(A, \circ)$ be a pre-Lie algebra and $(l, r,
V)$ be a bimodule. A linear map $T: V\rightarrow A$ is called an
{\it $\mathcal {O}$-operator associated to $(l, r, V)$} if $T$
satisfies
\begin{equation}
T(u)\circ T(v) = T(l(T(u))v + r(T(v))u), \forall u, v \in V.
\end{equation}
In particular, an ${\mathcal O}$-operator of a pre-Lie algebra
$(A, \circ)$ associated to the bimodule $(L_\circ, R_\circ, A)$ is
called a {\it Rota-Baxter operator (of weight zero) on
$(A,\circ)$}.

(3) (\cite{BLN})  Let $(A, \triangleright, \triangleleft)$ be an
L-dendriform algebra and $(l_\triangleright, r_\triangleright,
l_\triangleleft, r_\triangleleft, V)$ be a bimodule. A linear map
$T: V \rightarrow A$ is called an {\it $\mathcal {O}$-operator of
$(A, \triangleright, \triangleleft)$ associated to
$(l_\triangleright, r_\triangleright, l_\triangleleft,
r_\triangleleft, V)$} if $T$ satisfies (for any $u,v\in V$)
\begin{equation}
T(u) \triangleright T(v) =T[l_\triangleright(T(u))v +
r_\triangleright(T(v))u],
\end{equation}
\begin{equation}
T(u) \triangleleft T(v) = T[l_\triangleleft(T(u))v +
r_\triangleleft(T(v))u].
\end{equation}
In particular, an ${\mathcal O}$-operator of an L-dendriform
algebra $(A, \triangleright, \triangleleft)$ associated to the
bimodule
$(L_{\triangleright},R_{\triangleright},L_{\triangleleft},R_{\triangleleft},
A)$ is called a {\it Rota-Baxter operator (of weight zero) on $(A,
\triangleright, \triangleleft)$}.}\end{defn}

\begin{exam}{\rm (1) (\cite{Bai1})
Let $T : V \rightarrow \frak g $ be an $\mathcal {O}$-operator of a
Lie algebra $\frak g $ associated to a representation $(\rho, V )$.
Then there exists a pre-Lie algebra structure on $V$ given by
\begin{equation}
u \circ v = \rho(T(u))v, \forall u,v \in V.
\end{equation}
In particular, if $R:\frak g\rightarrow \frak g$ is a Rota-Baxter
operator (of weight zero) on a Lie algebra $\frak g$, then there
exists a pre-Lie algebra structure on $\frak g$ given by
(\cite{GS})
\begin{equation}
x\circ y=[R(x), y],\;\;\forall x,y\in \frak g.
\end{equation}

(2) (\cite{BLN})  Let $T:V\rightarrow A$ be an $\mathcal
{O}$-operator of a pre-Lie algebra $(A, \circ)$  associated to a
bimodule $(l, r, V)$. Then there exists an L-dendriform algebra
structure on $V$ given by
\begin{equation}
u \triangleright v = l(T(u))v,u \triangleleft v = -r(T(u))v, \forall
u, v\in V.
\end{equation}
In particular, if $R:A\rightarrow A$ is a Rota-Baxter operator (of
weight zero) on a pre-Lie algebra $(A,\circ)$, then there exists
an L-dendriform algebra structure on $A$ given by
\begin{equation}
x \triangleright y = R(x)\circ y,x \triangleleft y = -y\circ R(x),
\forall x, y \in A.
\end{equation}

(3) (\cite{BLN}) Let $\frak g $ be a Lie algebra and
 $\{R_1,R_2\}$ be a pair of commuting Rota-Baxter operators (of
 weight zero). Then there exists an L-dendriform algebra structure on $\frak g
$ given by
\begin{equation}
x\triangleright y = [R_1(R_2(x)), y],x\triangleleft y = [R_2(x),
R_1(y)],\forall x, y \in \frak g.
\end{equation}
 }
\end{exam}

\begin{remark}{\rm
Let $\frak g $ be a Lie algebra and $r \in \frak g \otimes \frak g
$. Recall that $r$ is a {\it solution of classical Yang-Baxter
equation (CYBE) in $\frak g $} if
\begin{equation}
[r_{12},r_{13}] + [r_{12},r_{23}] + [r_{13},r_{23}] = 0.
\end{equation}
When $r$ is skew-symmetric, $r$ is a solution of CYBE in $\frak g
$ if and only if $r$ is an $\mathcal {O}$-operator of $\frak g$
associated to $({\rm ad}^*,\frak g ^*)$ (\cite{Ku}). If, in
addition, there exists a nondegenerate invariant symmetric
bilinear form on $\frak g$, then $r$ is a solution of CYBE in
$\frak g $ if and only if $r$ regarded as a linear map from $\frak
g$ to $\frak g$ is a Rota-Baxter operator (of weight zero), that
is $r$ satisfies Eq. (2.20) (\cite{Se}). In this sense, Eq. (2.20)
is called the {\it operator form of the CYBE}. }
\end{remark}

\begin{prop} {\rm (\cite{BLN})} Let $(A,\triangleright,\triangleleft)$
be an L-dendriform algebra and $r \in A \otimes A$ be
skew-symmetric. Then $r$ is an $\mathcal {O}$-operator of
 $(A,\triangleright,\triangleleft)$ associated to
 $(L_{\triangleright}^* + L_{\triangleleft}^* -
R_{\triangleright}^* - R_{\triangleleft}^*,R_{\triangleright}^*,\
R_{\triangleright}^* - L_{\triangleleft}^*,-(R_{\triangleright}^* +
R_{\triangleleft}^*),A^*)$ if and only if $r$ satisfies the
following equation:
\begin{equation}
r_{23} \triangleleft r_{13} = r_{13} \circ r_{12} + r_{23} \bullet
r_{12}.\label{LD}
\end{equation}
Moreover, if, in addition, $r$ is invertible, then $r$ satisfies
Eq.~(\ref{LD}) if and only if the (nondegenerate) bilinear form
$\mathcal {B}$ induced by $r$ through Eq. (1.9) satisfies
\begin{equation}
\mathcal {B}(x \triangleleft y, z) =  -\mathcal {B}(y, z \circ x) +
\mathcal {B}(x, z \bullet y), \;\;\forall x, y, z \in A.
\end{equation}

\end{prop}

\begin{defn}
{\rm Let $(A, \triangleright, \triangleleft)$ be an L-dendriform
algebra. Let $r\in A \otimes A$. Eq.~ (\ref{LD}) is called {\it
$LD$-equation} in $(A, \triangleright, \triangleleft)$. On the
other hand, a skew-symmetric bilinear form $\mathcal {B}$ on $A$
satisfying Eq. (2.31) is called a {\it 2-cocycle} of  $(A,
\triangleright, \triangleleft)$.}
\end{defn}

\begin{remark}
{\rm In fact,  the $LD$-equation in an L-dendriform algebra is
regarded as an analogue of the CYBE in a Lie algebra (\cite{BLN},
also see Remark 2.13).}
\end{remark}

\section{L-quadri-algebras.} \setcounter{equation}{0}
\renewcommand{\theequation}
{3.\arabic{equation}}

\subsection{Definition and basic properties}

\begin{defn} {\rm Let $A$ be a vector space  with four bilinear products
$\searrow,  \nearrow,  \nwarrow$,  $\swarrow : A\otimes A\rightarrow
A$.  $(A, \searrow,  \nearrow,  \nwarrow,  \swarrow )$ is called an
{\it L-quadri-algebra} if for any $x, y, z \in A$,
\begin{equation}
x\searrow  (y\searrow z) -(x\ast y)\searrow z= y\searrow (x\searrow
z)- (y\ast x)\searrow z,\label{def-LQ-1}
\end{equation}
\begin{equation}
 x\searrow (y \nearrow z)- (x \vee y)\nearrow z
  =  y \nearrow (x\succ z) -(y \wedge
x ) \nearrow z,\label{def-LQ-2}
\end{equation}
\begin{equation}
x \searrow (y \nwarrow z) -(x \searrow y)\nwarrow z = y \nwarrow
(x\ast z)- (y \nwarrow x)\nwarrow z,\label{def-LQ-3}
 \end{equation}
\begin{equation}
x \nearrow (y\prec z)- ( x \nearrow y ) \nwarrow z = y \swarrow
(x\wedge z) - (y \swarrow x)\nwarrow z,\label{def-LQ-4}
\end{equation}
\begin{equation}
x \searrow (y \swarrow z)-(x \succ y)\swarrow z = y \swarrow (x\vee
z) -(y \prec x) \swarrow z,\label{def-LQ-5}
\end{equation}
where
\begin{equation}
x \succ y = x \searrow y + x \nearrow y,  x \prec y = x \nwarrow y +
x \swarrow y,\label{def-LD-v}
\end{equation}
\begin{equation}
x \vee y = x \searrow y + x \swarrow y,  x \wedge y = x \nearrow y +
x \nwarrow y,\label{def-LD-d}
\end{equation}
\begin{equation} x \ast y = x \searrow y + x \nearrow y
+ x \nwarrow y + x \swarrow y=x \succ y +x \prec y=x \vee y + x
\wedge y.\label{def-PL-sum}
\end{equation}
}\end{defn}

\begin{remark}{\rm Let $(A, \searrow,  \nearrow,  \nwarrow,  \swarrow )$ be an
L-quadri-algebra. If $\nearrow = \nwarrow = \swarrow = 0$, then
$(A,\searrow)$ is a pre-Lie algebra. If $\nwarrow = \searrow =
\swarrow =0$, or $\searrow = \nearrow = \swarrow =0$, or $\nearrow =
\nwarrow = \searrow = 0$ respectively, then $(A,\nearrow)$ or $(A,
\nwarrow)$ or $(A, \swarrow)$ is an associative algebra
respectively. For the cases that two operations among $\searrow,\
\nearrow,  \nwarrow,  \swarrow$ are zero, we have the following
results:

 (i) If $\nearrow = \swarrow =0$, then $(A,\searrow,
\nwarrow)$ is an L-dendriform algebra.

(ii) If $\nwarrow = \swarrow = 0$, then  $(A, \searrow ,\nearrow )$
is an L-dendriform algebra.

(iii) If $\nearrow = \nwarrow =0$, then $(A, \searrow, \swarrow)$ is
an L-dendriform algebra.

(iv) If $\searrow = \nearrow =0$, then $(A,\swarrow , \nwarrow )$ is
a dendriform algebra.

(v) If $\searrow =\swarrow =0$, then $(A, \nearrow , \nwarrow )$ is
a dendriform algebra.

(vi) If $\searrow = \nwarrow = 0$, then $$x\nearrow (y\swarrow z) =
y\swarrow (x\nearrow z),\; $$$$y\nearrow (x\nearrow z) = (y \nearrow
x-x\swarrow y)\nearrow z,\; y\swarrow (x\swarrow z)=(y\swarrow
x-x\nearrow y)\swarrow z.$$ Hence $(A, -\nearrow, \swarrow^\sigma)$
is a dendriform algebra, where $x\swarrow^\sigma y=y\swarrow x$ for
any $x,y\in A$.}\end{remark}

\begin{prop}\label{LQP}
Let $(A, \searrow,  \nearrow,  \nwarrow,  \swarrow )$ be an
L-quadri-algebra.

(1) The products given by Eq. (\ref{def-LD-v}) define an
L-dendriform algebra $(A, \succ, \prec)$ which is called the
associated $vertical$ L-dendriform algebra of $(A, \searrow,
\nearrow, \nwarrow,\swarrow )$

(2) The products given by Eq. (\ref{def-LD-d}) define
 an L-dendriform algebra $(A, \vee,
\wedge)$ which is called the associated $depth$ L-dendriform algebra
of $(A, \searrow,  \nearrow, \nwarrow, \swarrow )$.

(3) The products given
\begin{equation} x \triangleright y = x \searrow y
- y \nwarrow x,  x \triangleleft y = x \nearrow y - y \swarrow
x,\;\forall x,y\in A, \label{def-LD-h}
\end{equation}
define an L-dendriform algebra $(A,  \triangleright, \triangleleft)$
which is called the associated horizontal L-dendriform algebra of
$(A, \searrow,\nearrow, \nwarrow, \swarrow )$.
\end{prop}

\begin{proof} We only give an explicit proof of (3) since the proof of the other
two cases is similar. In this case, we show that
$$\begin{array}{rcl}
&&x \triangleright (y \triangleright z) = x \searrow (y \searrow z -
z \nwarrow y) - (y \searrow z - z \nwarrow y) \nwarrow x,\\
\\
&&(x \triangleright y) \triangleright z = (x \searrow y - y \nwarrow
x) \searrow z - z \nwarrow (x \searrow y - y \nwarrow x),\\
\\
&&(x \triangleleft y) \triangleright z = (x \nearrow y - y \swarrow
x) \searrow z - z \nwarrow (x \nearrow y - y \swarrow x),\\
\\
&&y \triangleright (x \triangleright z) = y \searrow (x \searrow z -
z \nwarrow x) - (x \searrow z - z \nwarrow x) \nwarrow y,\\
\\
&&(y \triangleright x) \triangleright z = (y \searrow x - x \nwarrow
y) \searrow z - z \nwarrow (y \searrow x - x \nwarrow y),\\
\\
&&(y \triangleleft x) \triangleright z = (y \nearrow x - x \swarrow
y) \searrow z - z \nwarrow (y \nearrow x - x \swarrow y).
\end{array}$$
By the definition of L-quadri-algebra, we show that
$$(x\triangleright y)\triangleright z +
(x\triangleleft y)\triangleright z + y\triangleright(x\triangleright
z) - (y\triangleleft x)\triangleright z - (y\triangleright
x)\triangleright z - x\triangleright(y\triangleright z)= 0.$$
Similarly we have
$$(x\triangleright y)\triangleleft z
+ y\triangleleft (x\triangleright z) + y\triangleleft(x\triangleleft
z) - (y\triangleleft x)\triangleleft z -
x\triangleright(y\triangleleft z) = 0.$$ Therefore $(A,
\triangleright, \triangleleft)$ is an L-dendriform algebra.
\end{proof}

\begin{coro} Let $(A, \searrow,  \nearrow,  \nwarrow,  \swarrow
)$ be an L-quadri-algebra and $(A,  \triangleright, \triangleleft)$,
$(A,  \succ, \prec)$, $(A,\ \vee, \wedge)$ be the associated
horizontal, vertical and depth L-dendriform algebras  respectively.
 Let $x, y \in A$.

(1)    If we define
\begin{equation} x \circ y = x \searrow y + x \swarrow y
- y \nwarrow x - y \nearrow x = x \triangleright y -y \triangleleft
x =x \vee y - y \wedge x,
\end{equation}
then $(A, \circ)$ is the associated vertical pre-Lie algebra of both
$(A, \triangleright, \triangleleft)$ and $(A,\ \vee, \wedge)$.

(2)    If we define (also see Eq. (3.8))
\begin{equation} x \ast y = x \searrow y + x \nearrow y
+ x \nwarrow y + x \swarrow y=x \succ y +x \prec y=x \vee y + x
\wedge y,
\end{equation}
then $(A, \ast)$ is the associated horizontal pre-Lie algebra of
both $(A,  \succ, \prec)$ and $(A,\vee, \wedge)$.

(3)    If we define
\begin{equation} x \bullet y = x \searrow y + x \nearrow y
- y \nwarrow x - y \swarrow x= x \triangleright y +x \triangleleft y
=x \succ y -y \prec x,
\end{equation}
then $(A, \bullet)$ is the associated horizontal pre-Lie algebra of
 $(A,  \triangleright,
\triangleleft)$ and the associated vertical pre-Lie algebra of $(A,
\succ, \prec)$.

(4) If we define
\begin{equation}
[x, y] = x \searrow y + x \swarrow y + x \nearrow y + x \nwarrow y -
(y \searrow x + y \swarrow x + y \nearrow x + y \nwarrow x),
\end{equation}
then $(A,[,])$ is a Lie algebra. $(A,[,])$ is called the
sub-adjacent Lie algebra of $(A, \searrow,  \nearrow,  \nwarrow,\
\swarrow )$ and we denote it by $\frak g(A)$.

\end{coro}

\begin{proof}
It follows from Proposition~\ref{LP} and Proposition~\ref{LQP}.
\end{proof}

\begin{prop} \label{def-module} Let $A$ be a vector
space with four bilinear products denoted by $\searrow, \nearrow$,
$\nwarrow, \swarrow : A\otimes A\rightarrow A$.

(1) $(A,  \searrow,  \nearrow,  \nwarrow,  \swarrow)$ is an
L-quadri-algebra if and only if  $(A,  \triangleright,
\triangleleft)$ defined by Eq. (\ref{def-LD-h}) is an L-dendriform
algebra and $(L_\searrow ,  -L_\nwarrow ,  L_\nearrow ,  -L_\swarrow
,  A)$ is a bimodule.

(2) $(A,  \searrow,  \nearrow,  \nwarrow,  \swarrow)$ is an
L-quadri-algebra if and only if  $(A,  \succ,  \prec)$ defined by
Eq. (\ref{def-LD-v}) is an L-dendriform algebra and $(L_\searrow ,
R_\nearrow ,  L_\swarrow ,  R_\nwarrow ,  A)$ is a bimodule.

(3) $(A,  \searrow,  \nearrow,  \nwarrow,  \swarrow)$ is an
L-quadri-algebra if and only if $(A,  \vee,\wedge)$ defined by Eq.
(\ref{def-LD-d}) is an L-dendriform algebra and $(L_\searrow ,
R_\swarrow , L_\nearrow ,  R_\nwarrow ,  A)$ is a
bimodule.\end{prop}

\begin{proof}
The conclusions can be obtained by a straightforward computation or
a similar proof as of Proposition 3.11.
\end{proof}

\begin{remark}
{\rm In the sense of the above conclusion (1), an L-quadri-algebra
is understood as an algebra structure whose left multiplications
give a bimodule structure on the underlying vector space of the
L-dendriform algebra defined by certain commutators. It can be
regarded as the ``rule" of introducing the notion of
L-quadri-algebra, which more generally, is the ``rule" of
introducing the notions of the Loday algebras and their Lie
algebraic analogues (\cite{Bai3}, also see Proposition 2.3 and
Proposition 2.8). }
\end{remark}

\begin{coro}
Let $(A,  \searrow,  \nearrow,  \nwarrow\ \swarrow)$ be an
L-quadri-algebra and $(A, \triangleright, \triangleleft)$, $(A,
\succ, \prec)$, $(A,\ \vee, \wedge)$ be the associated horizontal,
vertical and depth L-dendriform algebras of $(A, \searrow, \nearrow,
\nwarrow, \swarrow )$ respectively.

(1) $(L_\ast^*, -L_\nwarrow^*, -L_\wedge^*, L_\prec^*,  A^*)$ is a
bimodule of $(A, \triangleright , \triangleleft )$.

(2)  $(L_\circ^*, R_\nearrow^*, L_\triangleright^*, -R_\wedge^*,
A^*)$ is a bimodule of $(A, \succ, \prec)$.

(3)  $(L_\bullet^*, R_\swarrow^*, -L_\triangleleft^*, -R_\prec^*,
A^*)$ is a bimodule of  $(A, \vee, \wedge)$.
\end{coro}

\begin{proof}
It follows from Proposition 2.10 and Proposition 3.5.
\end{proof}

\begin{prop} Let $(A,\searrow , \nearrow , \nwarrow ,\swarrow )$ be an
L-quadri-algebra.

(1) [\textbf{Transpose of an L-quadri-algebra}]\;\;  Define four
bilinear products $\searrow^t,  \nearrow^t,  \nwarrow^t,\
\swarrow^t$ on $A$ by
\begin{equation}
x\searrow^t y=x\searrow y,\;\; x\nwarrow^t y=x\nwarrow y,\;\;
x\nearrow^t y = x\swarrow y,\;\; x\swarrow^t y= x\nearrow y,\;\;
\forall x,y\in A.
\end{equation}
Then $(A,\searrow^t,  \nearrow^t,  \nwarrow^t,  \swarrow^t)$ is an
L-quadri-algebra which is called the transpose of $(A,\searrow ,
\nearrow , \nwarrow$, $\swarrow )$. Moreover (for any $x, y\in A$)
\begin{equation}
x\vee^t y=  x\succ y,\;\;x\wedge^t y= x\prec y, x\triangleright^t y=
x\triangleright y , x\triangleleft^t y= -y\triangleleft x, x\succ^t
y=x \vee y, x \prec^t y =x \wedge y.
\end{equation}

(2) [\textbf{Symmetry} $Sym^a$]   Define four bilinear products
$\searrow^a,  \nearrow^a,  \nwarrow^a,  \swarrow^a$ on $A$ by
\begin{equation}
x\searrow^a y=x\searrow y,  x\nwarrow^a y=-y\nearrow x, x\nearrow^a
y = -y\nwarrow  x,  x\swarrow^a y= x\swarrow y,  \forall x,y\in A.
\end{equation}
Then $(A,\searrow^a,  \nearrow^a,  \nwarrow^a,  \swarrow^a)$ is an
L-quadri-algebra and (for any $x, y\in A$)
\begin{equation}
x \vee^a y=  x \vee y, x\wedge^a y= -y\wedge x, x\triangleright^a y
= x\succ y, x\triangleleft^a y= -y\prec x, x \succ^a y=
x\triangleright y, x\prec^a y =-y\triangleleft x.
\end{equation}

(3) [\textbf{Symmetry} $Sym^b$]  Define four bilinear products
$\searrow^b,  \nearrow^b,  \nwarrow^b,  \swarrow^b$ on $A$ by
\begin{equation}
x\searrow^b y=x\searrow y,  x\nwarrow^b y=-y\nearrow x, x\nearrow^b
y = x\swarrow y,  x\swarrow^b y= -y\nwarrow x,  \forall x,y\in A.
\end{equation}
Then $(A,\searrow^b,  \nearrow^b,  \nwarrow^b,  \swarrow^b)$ is an
L-quadri-algebra and (for any $x, y\in A$)
\begin{equation}
x\vee^b y=  x \triangleright y, x\wedge^b y=-y \triangleleft x,
x\triangleright^b y= x\succ y, x\triangleleft^b y= x\prec y, x
\succ^b y=x\vee y, x\prec^b y=-y\wedge x.
\end{equation}

(3) [\textbf{Symmetry} $Sym^c$]  Define four bilinear products
$\searrow^c,  \nearrow^c,  \nwarrow^c,  \swarrow^c$ on $A$ by
\begin{equation}
x\searrow^c y=x\searrow y,  x\nwarrow^c y=-y\swarrow x, x\nearrow^c
y = -y\nwarrow x,  x\swarrow^c y= x\nearrow y,  \forall x,y\in A.
\end{equation}
Then $(A,\searrow^c,  \nearrow^c,  \nwarrow^c,  \swarrow^c)$ is an
L-quadri-algebra and (for any $x, y\in A$)
\begin{equation}
x\vee^c y=  x\succ y,x\wedge^c y=-y \prec x, x\triangleright^c y=
x\vee y , x\triangleleft^c y= -y \wedge x , x\succ^c
y=x\triangleright y, x\prec^c y=x \triangleleft y.
\end{equation}

(4) [\textbf{Symmetry} $Sym^d$]  Define four bilinear products
$\searrow^d,  \nearrow^d,  \nwarrow^d,  \swarrow^d$ on $A$ by
\begin{equation}
x\searrow^d y=x\searrow y,  x\nwarrow^d y=-y\swarrow x, x\nearrow^d
y = x\nearrow y,  x\swarrow^d y= -y\nwarrow x,  \forall x,y\in A.
\end{equation}
Then $(A,\searrow^d,  \nearrow^d,  \nwarrow^d,  \swarrow^d)$ is an
L-quadri-algebra and (for any $x, y\in A$)
\begin{equation}
x\vee^d y=  x\triangleright y, x\wedge^d y=x\triangleleft y,
x\triangleright^d y= x\vee y , x\triangleleft^d y= x \wedge y ,
x\succ^b y=x\succ y, x\prec^d y=-y\prec x.
\end{equation}
\end{prop}

\begin{proof}

In fact, there are 5 non-trivial symmetries in the axioms defining
an L-quadri-algebra which are consistent with the symmetry group
$\Sigma_3$. By a straightforward checking, the above 5 cases
correspond to these symmetries respectively.
\end{proof}

\begin{lemma} {\rm (\cite{BLN})}
Let $(l_{\triangleright}, r_{\triangleright}, l_{\triangleleft},
r_{\triangleleft}, V)$ be a bimodule of an L-dendriform algebra
$(A,\triangleright, \triangleleft)$.

(1) $(l_{\triangleright},-l_{\triangleleft},V)$ is a bimodule of the
associated vertical pre-Lie algebra $(A,\circ)$.

(2) $(l_{\triangleright},r_{\triangleleft},V)$ is a bimodule of the
associated horizontal pre-Lie algebra $(A,\bullet)$.

(3) $(l_{\triangleright},V)$ is a representation of the sub-adjacent
Lie algebra $\frak g  (A)$. \label{LD-module}
\end{lemma}

\begin{coro}
Let $(A, \searrow,  \nearrow,  \nwarrow,  \swarrow )$ be an
L-quadri-algebra.

(1)    $(L_\searrow ,  -L_\nearrow ,  A)$ is a bimodule of the
associated pre-Lie algebra $(A, \circ)$.

(2)    $(L_\searrow,  R_\nwarrow ,  A)$ is a bimodule of the
associated pre-Lie algebra $(A, \ast)$.

(3)    $(L_\searrow ,  -L_\swarrow ,  A)$ is a bimodule of the
associated pre-Lie algebra $(A, \bullet)$.

(4)\quad $(L_\searrow,A)$ is a representation of the sub-adjacent
Lie algebra $\frak g (A)$.\end{coro}

\begin{proof}

It follows from Lemma~\ref{LD-module}, Corollary 3.4 and
Proposition~\ref{def-module}. \end{proof}

\subsection{Constructions of L-quadri-algebras}

For brevity, in this subsection, we only give the study involving
the associated horizontal L-dendriform algebras. The corresponding
study on the other associated L-dendriform algebras can be obtained
by the transpose and symmetries of the L-dendriform algebras through
Proposition 3.8.

\begin{prop} Let $(l_{\triangleright},r_{\triangleright},l_{\triangleleft},r_{\triangleleft},
V)$ be a bimodule of an L-dendriform algebra
 $(A, \triangleright, \triangleleft)$. If $T$ is
an $\mathcal {O}$-operator associated to
$(l_{\triangleright},r_{\triangleright},l_{\triangleleft},r_{\triangleleft},
V)$, then there exists an L-quadri-algebra structure on $V$
defined by (for any $u,v\in V$)
\begin{equation}
u \searrow v = l_\triangleright (T(u))v,  u \nwarrow v =
-r_\triangleright(T(u))v,  u \nearrow v = l_\triangleleft(T(u))v,\
u\swarrow v = -r_\triangleleft(T(u))v.
\end{equation}
Therefore there is an L-dendriform algebra structure on $V$ defined
by Eq. (3.9) as the associated horizontal L-dendriform algebra of
$(V,\searrow , \nearrow , \nwarrow ,\swarrow )$ and $T$ is a
homomorphism of L-dendriform algebras. Furthermore, $T(V) =
\{T(v)\mid v \in V\} \subset A$ is an L-dendriform subalgebra of
$(A, \triangleright, \triangleleft)$ and there is an induced
L-quadri-algebra structure on $A$ given by
\begin{equation}
\begin{array}{rcl}
&&T(u) \searrow T(v) = T(u \searrow v),  T(u) \nearrow T(v) = T(u
\nearrow v),\\
\\
&&T(u) \nwarrow T(v) = T(u \nwarrow v),  T(u) \swarrow T(v) = T(u
\swarrow v), \forall\; u, v \in V.
\end{array}
\end{equation}
Moreover, the corresponding associated horizontal L-dendriform
algebra structure on $T(V)$ is just the L-dendriform algebra
structure of $(A,\triangleright, \triangleleft)$ and $T$ is a
homomorphism of L-quadri-algebras.
\end{prop}

\begin{proof} By Eq. (3.24), for any $u, v, w \in V$, we show that
\begin{eqnarray*}
 &&u\searrow (v \searrow w) - v\searrow (u \searrow w) =
[l_\triangleright(T(u)), l_\triangleright (T(v))]w = l_\triangleright[T(u), T(v)]w,\\
&& [u, v] \searrow w = T(l_\triangleright(T(u))v +
T(r_\triangleright(T(v))u) + T(l_\triangleleft(T(u))v +
T(r_\triangleleft(T(v))u)\\
 &&\hspace{2.8cm} - T(l_\triangleright(T(v))u +
T(r_\triangleright(T(u))v) - T(l_\triangleleft(T(v))u +
T(r_\triangleleft(T(u))v)\\
&&\hspace{2cm}=\ l_\triangleright [T(u), T(v)]w.
\end{eqnarray*}
Therefore we show that $ u\searrow (v \searrow w) - v\searrow (u
\searrow w) = [u, v] \searrow w$. By a similar way, we show that
Eqs. (3.2)-(3.5) holds for $(V, \searrow,  \nearrow,  \nwarrow,
\swarrow)$. The other conclusions follow immediately.
\end{proof}

\begin{coro} Let  $R$ be a Rota-Baxter operator of (weight zero) on an L-dendriform algebra
 $(A, \triangleright, \triangleleft)$. Then there exists an L-quadri-algebra structure on $A$ defined
by
\begin{equation}
x \searrow y = R(x)\triangleright y,  x \nwarrow y = - y
\triangleright R(x),  x \nearrow y = R(x)\triangleleft y,  x\swarrow
y= -y \triangleleft R(x), \forall x, y \in A.
\end{equation}
Its transpose is given by
\begin{equation}
x \searrow^t y = R(x)\triangleright y,  x \nwarrow^t y = - y
\triangleright R(x),  x \nearrow^t y = -y \triangleleft R(x),\
x\swarrow^t y= R(x)\triangleleft y, \forall x, y \in A.
\end{equation}
Its symmetries are given by (for any $x,y\in A$)
\begin{equation}
x \searrow^a y = R(x)\triangleright y,  x \nwarrow^a y = - R(y)
\triangleleft x,  x \nearrow^a y = x\triangleright R(y),
x\swarrow^a y= -y\triangleleft R(x);
\end{equation}
\begin{equation}
x \searrow^b y = R(x)\triangleright y,  x \nwarrow^b y = - R(y)
\triangleleft x,  x \nearrow^b y = -y\triangleleft R(x),
x\swarrow^b y= x\triangleright R(y);
\end{equation}
\begin{equation}
x \searrow^c y = R(x)\triangleright y,  x \nwarrow^c y = x
\triangleleft R(y),  x \nearrow^c y = x\triangleright R(y),\
x\swarrow^c y= R(x)\triangleleft y;
\end{equation}
\begin{equation}
x \searrow^d y = R(x)\triangleright y,  x \nwarrow^d y = x
\triangleleft R(y),  x \nearrow^d y = R(x)\triangleleft y,\
x\swarrow^d y= x\triangleright R(y).
\end{equation}
\end{coro}

\begin{proof}

It follows from Proposition 3.11 by choosing the bimodule
$(L_{\triangleright},R_{\triangleright},L_{\triangleleft},R_{\triangleleft},
A)$.\end{proof}

\begin{lemma}
 Let $\{R_1,R_2\}$ be a pair of commuting Rota-Baxter
operators (of weight zero) on a pre-Lie algebra $(A, \circ)$. Then
$R_2$ is a Rota-Baxter operator (of weight zero) on the L-dendriform
algebra defined by Eq. (2.27) with $R=R_1$.\end{lemma}

\begin{proof}
For any $x,y\in A$, we show that
\begin{eqnarray*}
&&R_2(x)\triangleright R_2(y)=R_1(R_2(x))\circ
R_2(y)=R_2(R_1(x))\circ R_2(y)\\
&&\hspace{2.5cm} = R_2(R_2 R_1(x)\circ y+R_1(x)\circ
R_2(y))=R_2(R_2(x)\triangleright
y+x\triangleright R_2(y));\\
&&R_2(x)\triangleleft R_2(y)=-R_2(y) \circ R_1(R_2(x))=-R_2(y)\circ
R_2(R_1(x))\\
&&\hspace{2.5cm} =-R_2(R_2(y)\circ R_1(x)+y\circ
R_2R_1(x))=R_2(x\triangleleft R_2(y)+R_2(x)\triangleleft y).
\end{eqnarray*}
Note that the above identities involve the axioms of a Rota-Baxter
operator (of weight zero) and $R_1R_2=R_2R_1$. Hence $R_2$ is a
Rota-Baxter operator (of weight zero) on $(A,\triangleright,
\triangleleft)$.
\end{proof}

\begin{coro} Let $\{R_1,R_2\}$ be a pair of commuting Rota-Baxter
operators (of weight zero) on a pre-Lie algebra $(A, \circ)$. Then
there exists an L-quadri-algebra structure on $A$ defined by (for
any $x, y \in A$)
\begin{equation}
x\searrow y = R_1R_2(x) \circ y,  x\nwarrow y = -R_1(y)\circ R_2(x),
x\nearrow y = -y \circ R_1R_2(x),  x\swarrow y = R_2(x)\circ R_1(y).
\end{equation}
\end{coro}

\begin{proof}
It follows from Corollary 3.11 and Lemma 3.13.
\end{proof}

By a similar proof as of Lemma 3.13, we have the following
conclusion.

\begin{lemma}
Let
 $\{R_1,R_2, R_3\}$ be three pairwise commuting Rota-Baxter
 operators (of weight zero) of a Lie algebra $\frak g $. Then $R_3$ is a Rota-Baxter operator (of weight zero) on the
L-dendriform algebra defined by Eq. (2.28).
\end{lemma}

\begin{coro} Let
 $\{R_1,R_2, R_3\}$ be three pairwise commuting Rota-Baxter
 operators of a Lie algebra $\frak g $. Then there exists an L-quadri-algebra structure on
$\frak g $ defined by
\begin{equation}
\begin{array}{rcl}
&&x\searrow y = [R_1(R_2(R_3(x))), y],\;\;  x\nwarrow y = [R_3(x),
R_1(R_2(y))],  \\
\\
&&x \nearrow y = [R_2(R_3(x)), R_1(y)],\;\; x\swarrow y =
[R_1(R_3(x)), R_2(y)], \;\;\forall x, y \in \frak g .
\end{array}
\end{equation}
\end{coro}

\begin{proof}
It follows from Corollary 3.12 and Lemma 3.15.
\end{proof}

\begin{theorem} Let $(A, \triangleright, \triangleleft)$ be an L-dendriform
algebra. Then there exists an L-quadri-algebra structure on $A$
such that $(A, \triangleright, \triangleleft)$ is the associated
horizontal L-dendriform algebra if and only if there exists an
invertible $\mathcal {O}$-operator of $(A, \triangleright,
\triangleleft)$.
\end{theorem}

\begin{proof}
Suppose that there exists an invertible ${\mathcal O}$-operator
$T$ of $(A, \triangleright, \triangleleft)$ associated to a
bimodule
$(l_{\triangleright},r_{\triangleright},l_{\triangleleft},r_{\triangleleft},
V)$. By Proposition 3.11, there exists an L-quadri-algebra
structure on $V$ given by Eq. (3.24). Therefore we define an
L-quadri-algebra structure on $A$ by Eq. (3.25) such that $T$ is
an isomorphism of L-quadri-algebras, that is,
\begin{eqnarray*}
&&x \searrow y = T(l_\triangleright (x) T^{-1}(y)),  x \nwarrow y =
-T(r_\triangleright(x)T^{-1}(y)),\\
&&  x \nearrow y = T(l_\triangleleft(x)T^{-1}(y)), x\swarrow y =
-T(r_\triangleleft(x) T^{-1}(y)),\;\;\forall x,y\in A.
\end{eqnarray*}
Moreover, the associated horizontal L-dendriform algebra is $(A,
\triangleright, \triangleleft)$ since (for any $x,y\in A$)
\begin{eqnarray*}
&&x \searrow y - y \nwarrow x =T(l_\triangleright (x) T^{-1}(y))+
T(r_\triangleright(y)T^{-1}(x)) =T(T^{-1}(x))\triangleright
T(T^{-1}(y))=x\triangleright y,\\
&&x \nearrow y - y \swarrow x =T(l_\triangleleft (x) T^{-1}(y))+
T(r_\triangleleft(y)T^{-1}(x)) =T(T^{-1}(x))\triangleleft
T(T^{-1}(y))=x\triangleleft y.
\end{eqnarray*}
 Conversely, let $(A, \searrow, \nearrow, \nwarrow, \swarrow)$  be an
L-quadri-algebra and $(A, \triangleright, \triangleleft)$ be the
associated horizontal L-dendriform algebra. Then the identity map
${\rm id}: A\rightarrow A$ is an $\mathcal {O}$-operator of $(A,
\triangleright, \triangleleft)$ associated to the bimodule
$(L_\searrow ,  -L_\nwarrow ,  L_\nearrow ,  -L_\swarrow ,  A)$.
\end{proof}

The following conclusion reveals {\rm the relationship between  the
L-dendriform algebras with a nondegenerate (skew-symmetric)
2-cocycle and L-quadri-algebras}:

\begin{prop}
Let $(A, \triangleright, \triangleleft)$ be an L-dendriform
algebra with a nondegenerate 2-cocycle $\mathcal {B}$. Then there
exists an L-quadri-algebra structure $(A, \searrow, \nearrow,
\nwarrow,  \swarrow)$ on $A$ defined by
\begin{equation}
\begin{array}{rcl}
&&\mathcal {B} (x \searrow y, z) = - \mathcal {B}(y, [x,z]),\
\mathcal {B}(x \nwarrow y, z) = \mathcal {B} (y, z \triangleright x),\\
\\
&&\mathcal {B} (x \nearrow y, z) = -\mathcal {B} (y,  z \circ x),\
\mathcal {B} (x \swarrow y, z) = - \mathcal {B} (y, z \bullet x),
\forall x, y, z \in A,
\end{array}
\end{equation}
such that $(A, \triangleright, \triangleleft)$ is the associated
horizontal L-dendriform algebra.
\end{prop}

\begin{proof}
By Proposition 2.14, the invertible linear map $T:A^*\rightarrow A$
defined by  Eq. (1.9) is an invertible $\mathcal O$-operator of $(A,
\triangleright, \triangleleft)$ associated to the bimodule
$(L_{\triangleright}^* + L_{\triangleleft}^* - R_{\triangleright}^*
- R_{\triangleleft}^*,R_{\triangleright}^*, R_{\triangleright}^* -
L_{\triangleleft}^*,
 -(R_{\triangleright}^* +
R_{\triangleleft}^*),\ A^*)$. By Theorem 3.17, there is an
L-quadri-algebra structure on $A$ defined by
\begin{eqnarray*} \mathcal {B}(x\searrow y, z) &=&
\mathcal {B}(T(L_{\triangleright}^* + L_{\triangleleft}^* -
R_{\triangleright}^* - R_{\triangleleft}^*)(x)T^{-1}(y)), z) =
\langle (L_{\triangleright}^* + L_{\triangleleft}^* -
R_{\triangleright}^* - R_{\triangleleft}^*)(x)T^{-1}(y), z\rangle
\\
&=& - \langle T^{-1}(y), [x, z]
\rangle  = - \mathcal {B}(y, [x,z]);\\
\mathcal {B}(x\nwarrow y,z)&=&\mathcal
B(T(R_{\triangleright}^*(x)T^{-1}(y)),z)= \langle
R_{\triangleright}^*(x)T^{-1}(y),z\rangle=-\langle T^{-1}(y),
z\triangleright x\rangle=
 -\mathcal {B}(y,z\triangleright x);\\
\mathcal {B}(x\nearrow y, z) &=& \mathcal {B}(T(R_{\triangleright}^*
- L_{\triangleleft}^*)(x)T^{-1}(y)), z) = 
=- \langle T^{-1}(y), z\triangleright x-x\triangleleft z
\rangle  = - \mathcal {B}(y,z\circ x );\\
\mathcal {B}(x\swarrow y, z) &=& \mathcal
{B}(T(-R_{\triangleright}^* -R_{\triangleleft}^*)(x)T^{-1}(y)), z) 
=\langle T^{-1}(y), z\triangleright x+z\triangleleft x \rangle  =
\mathcal {B}(y,z\bullet x ),
\end{eqnarray*}
for any $x,y,z\in A$, such that $(A, \triangleright, \triangleleft)$
is the associated horizontal L-dendriform algebra.
\end{proof}

\begin{remark}
{\rm By Proposition 3.8, it is easy to get the transpose and
symmetries of the L-quadri-algebras in Corollary 3.14, Corollary
3.16, Theorem 3.17, and Corollary 3.18, like Corollary
3.12.}\end{remark}

\begin{coro}
Let $(A, \triangleright, \triangleleft)$ be an L-dendriform algebra
with a nondegenerate 2-cocycle $\mathcal {B}$. Then there exist two
other L-dendriform algebras $(A, \vee,\wedge)$ and $(A,\succ,
\prec)$ given by
\begin{equation}
\mathcal {B}(x \vee y, z) = -\mathcal {B} (y, x\ast z),  \mathcal
{B}(y \wedge x, z) = \mathcal {B} (x, y\triangleleft z), \forall
x, y, z \in A,
\end{equation}
\begin{equation}
\mathcal {B}(x \succ y, z) = -\mathcal {B} (y, x\circ z),\mathcal
{B}(y \prec x, z) = -\mathcal {B} (x, z\triangleleft y), \forall x,
y, z \in A,
\end{equation}
respectively. Moreover, they are the associated depth and vertical
L-dendriform algebras of the L-quadri-algebra $(A,\searrow,\
\nearrow,  \nwarrow,  \swarrow)$ given Eq. (3.34) respectively.
\end{coro}

\begin{proof} It follows from Proposition 3.3 and Proposition 3.18.  \end{proof}

The following conclusion provides {a construction of solutions
of $LD$-equation in certain L-dendriform algebras from
L-quadri-algebras}:

\begin{lemma} {\rm (\cite{BLN})} Let $(A, \triangleright, \triangleleft)$
be an L-dendriform algebra and $(l_\triangleright, r_\triangleright,
l_\triangleleft, r_\triangleleft, V)$ be a bimodule. Let
$(l_{\triangleright}^*+l_{\triangleleft}^*-r_{\triangleright}^*-r_{\triangleleft}^*,\
r_{\triangleright}^*,r_{\triangleright}^*-l_{\triangleleft}^*,\
-(r_{\triangleright}^*+r_{\triangleleft}^*),V^*)$ be the bimodule
given in Proposition 2.10. Let $T: V\rightarrow A$ be a linear map
which can be identified as an element in the vector space $(A\oplus
V)\otimes (A\oplus V)$.  Then $ r = T - \sigma(T)$  is a
skew-symmetric solution of $LD$-equation in the L-dendriform algebra
$A\ltimes_{l_{\triangleright}^*+l_{\triangleleft}^*-r_{\triangleright}^*-r_{\triangleleft}^*,\
r_{\triangleright}^*,r_{\triangleright}^*-l_{\triangleleft}^*,\
-(r_{\triangleright}^*+r_{\triangleleft}^*)}V^*$ if and only if $T$
is an $\mathcal{O}$-operator of  $(A, \triangleright,
\triangleleft)$ associated to $(l_\triangleright, r_\triangleright,
l_\triangleleft, r_\triangleleft, V)$.
\end{lemma}

\begin{coro}
Let $(A, \searrow,  \nearrow,  \nwarrow,  \swarrow)$ be an
L-quadri-algebra and $(A, \triangleright, \triangleleft)$,
$(A,\succ,  \prec)$, $(A, \vee,\wedge)$ be the associated
horizontal, vertical, and depth L-dendriform algebras
respectively. Let $\{e_1,\cdots$, $e_n\}$ be a basis of $A$ and
$\{e_1^*,..., e_n^*\}$ be the dual basis. Then
\begin{equation} r = \sum_{i=1}^n (e_i\otimes e_i^* - e_i^*\otimes
e_i)
\end{equation} is a skew-symmetric solution of $LD$-equation in the
L-dendriform algebras $$A\ltimes_{L_\ast^*, -L_\nwarrow^*,
-L_\wedge^*, L_\prec^*}A^*,\;\; A\ltimes_{L_\circ^*, R_\nearrow^*,
L_\triangleright^*, -R_\wedge^*}A^*,\;\;A\ltimes_{L_\bullet^*,
R_\swarrow^*, -L_\triangleleft^*, -R_\prec^*}A^*,$$ respectively.
Moreover there is a natural 2-cocycle $\mathcal {B}$ of these
L-dendriform algebras induced by $r$ through Eq. (1.9) which is
given by
\begin{equation}
{\mathcal B}(x+a^*,y+b^*)=-\langle a^*,y\rangle  +\langle
x,b^*\rangle  ,\;\; \forall x,y\in A,a^*,b^*\in A^*.
\end{equation}
\end{coro}

\begin{proof} Since  ${\rm id}$ is an $\mathcal {O}$-operator of
 the L-dendriform algebra $(A,  \triangleright,
\triangleleft)$ associated to the bimodule  $(L_\searrow ,
-L_\nwarrow , L_\nearrow , -L_\swarrow ,  A)$, the L-dendriform
algebra $(A,  \succ,  \prec)$ associated to the bimodule
$(L_\searrow , R_\nearrow , L_\swarrow , R_\nwarrow ,  A)$ and the
L-dendriform algebra $(A,  \vee,\wedge)$ associated to the
bimodule $(L_\searrow , R_\swarrow,$ $L_\nearrow , R_\nwarrow ,
A)$, the conclusion follows from Lemma 3.21.
\end{proof}

\subsection{Relationships with Loday algebras}

\begin{defn} {\rm (\cite{AL})\quad Let $A$ be a vector space  with four bilinear products
$\searrow,  \nearrow,  \nwarrow$,  $\swarrow : A\otimes
A\rightarrow A$.  $(A, \searrow,  \nearrow,  \nwarrow,  \swarrow
)$ is called a {\it quadri-algebra} if both of two sides of Eqs.
(\ref{def-LQ-1})-(\ref{def-LQ-5}) are zero.}
\end{defn}

\begin{remark}{\rm
Note that in the case of quadri-algebras, Eq. (\ref{def-LQ-1})
only gives an independent identity due to the symmetry in $x,y$,
whereas every other equation (Eqs.
(\ref{def-LQ-2})-(\ref{def-LQ-5})) provides 2 independent
identities. Therefore there are  9 independent identities for a
quadri-algebra. As a direct consequence, any quadri-algebra is an
L-quadri-algebra. }
\end{remark}

\begin{remark}
{\rm In fact, the both sides of Eqs. (2.5)-(2.6) and Eqs.
(3.1)-(3.5) can be regarded as the ``generalized associators". In
this sense, pre-Lie algebras, L-dendriform algebras and
L-quadri-algebras satisfy that the associators are left-symmetric
or the ``generalized associators" are ``generalized
left-symmetric", whereas associative algebras, dendriform algebras
and quadri-algebras satisfy that the associators or the
`generalized associators" are zero respectively.}
\end{remark}

\begin{prop}
Let $(A,\triangleright, \triangleleft)$ be a dendriform algebra.

(1) {\rm (\cite{Lo1})} The products given by Eq. (2.7) define an
associative algebra $(A,\bullet)$.

(2) {\rm (\cite{Ron},\cite{Cha})} The products given by Eq. (2.8)
define a pre-Lie algebra $(A,\circ)$.
\end{prop}

\begin{prop} Let $(A, \searrow, \nearrow, \nwarrow,
\swarrow)$ be a quadri-algebra.

(1) {\rm (\cite{AL})} The products given by Eq. (3.6) define a
dendriform algebra $(A,\succ,\prec)$.

(2) {\rm (\cite{AL})}  The products given by Eq. (3.7)
 define a dendriform algebra $(A,\vee,\wedge)$.

(3) {\rm (\cite{BLN})}  The products given by Eq. (3.9)
 define an L-dendriform algebra $(A,\triangleright,\triangleleft)$.

(4) {\rm (\cite{AL})} The products given by  Eq. (3.8) define an
associative algebra $(A,*)$.

(5) {\rm (\cite{BLN})}  The products given by Eq. (3.10) define a
pre-Lie algebra $(A,\circ)$.

(6) {\rm (\cite{BLN})} The products given by Eq. (3.12) define a
pre-Lie algebra $(A,\bullet)$.

(7) {\rm (\cite{BLN})}  The products given by Eq. (3.13) define a
Lie algebra $(\frak g(A),[,])$.
\end{prop}

The notion of {\it octo-algebra} was introduced in \cite{Le3} as a
Loday algebra with 8 operations. In order to get the explicit
relationships between octo-algebras and L-quadri-algebras, we need
put both of them into a bigger framework. So we introduce a notion
of {\it L-octo-algebra} as a natural generalization, which was put
in the Appendix due to the complexity.

\begin{defn}{\rm (\cite{Le3}) With notations in the Appendix,
let $A$ be a vector space over with eight bilinear products
denoted by $ \searrow_1, \searrow_2,\nearrow_1, \nearrow_2,
\nwarrow_1, \nwarrow_2, \swarrow_1, \swarrow_2: A\otimes A
\longrightarrow A$.  $(A, \searrow_1, \searrow_2$, $\nearrow_1,
\nearrow_2, \nwarrow_1, \nwarrow_2, \swarrow_1, \swarrow_2)$ is
called an {\it octo-algebra} if both sides of Eqs. (A1)-(A14)  are
zero. }
\end{defn}

\begin{coro} Any octo-algebra is an L-octo-algebra.\end{coro}

The following conclusion gives the explicit relationships between
octo-algebras and L-quadri-algebras (see Proposition A2).

 \begin{prop} With the notations in the Appendix. Let $(A, \searrow_1, \searrow_2,
 \nearrow_1, \nearrow_2, \nwarrow_1, \nwarrow_2, \swarrow_1$, $
\swarrow_2)$ be an octo-algebra.

(1) {\rm (\cite{Le3})}   $(A,  \vee_2, \wedge_2,\wedge_1, \vee_1)$
is a quadri-algebra.

(2) {\rm (\cite{Le3})}  $(A,  \succ_2, \succ_1,  \prec_1,\prec_2)$
is a quadri-algebra.

(3) {\rm (\cite{Le3})}  $(A,  \searrow_{12}, \nearrow_{12},
\nwarrow_{12},\ \swarrow_{12})$ is a quadri-algebra.

(4) $(A,  \triangleright_1^2, \triangleleft_1^2,\ \triangleleft_2^1,
\triangleright_2^1)$ is an L-quadri-algebra.
\end{prop}

Summarizing the above study in this subsection, we have the
following commutative diagram: (we omit ``algebra" for abbreviation)

$$\begin{matrix} \mbox{Lie }
&\stackrel{-}{\leftarrow} & \mbox{Pre-Lie}&
\stackrel{-,+}{\leftarrow} & \mbox{L-dendriform
}&\stackrel{-,+}{\leftarrow} & \mbox{L-quadri}&
\stackrel{-,+}{\leftarrow} & \mbox{L-octo}\cr &
\stackrel{-}{\nwarrow} & \Uparrow\in &\stackrel{-}{\nwarrow} &
\Uparrow \in &\stackrel{-}{\nwarrow} & \Uparrow\in
&\stackrel{-}{\nwarrow} & \Uparrow\in\cr
 & & \mbox{Associative } &\stackrel{+}{\leftarrow}
& \mbox{Dendriform }&\stackrel{+}{\leftarrow} & \mbox{Quadri}&
\stackrel{+}{\leftarrow} & \mbox{Octo}\cr
\end{matrix}$$
\noindent where $``\Uparrow\in "$ means the inclusion. $``+"$
means the operation $x\circ_1y+x\circ_2 y$ and  $``-"$ means the
operation $x\circ_1y-y\circ_2 x$.

\subsection{Bilinear forms on L-quadri-algebras and $LQ$-equation.}

\begin{defn} {\rm Let $(A, \searrow,  \nearrow,  \nwarrow,  \swarrow
)$ be an L-quadri-algebra. A skew-symmetric bilinear form
$\mathcal {B}$ on $A$ is called {\it invariant} if $\mathcal {B}$
satisfies Eq. (3.34). }\end{defn}

The following conclusion is obvious:

\begin{prop}
Let $(A, \nearrow,  \nwarrow,  \searrow,  \swarrow )$ be an
L-quadri-algebra and $\mathcal B$ be a skew-symmetric bilinear
form. If $\mathcal B$ is invariant on $A$, then $\mathcal B$ is a
2-cocycle on the associated horizontal, vertical and depth
L-dendriform algebras. Conversely, if $\mathcal B$ is a
nondegenerate (skew-symmetric) 2-cocycle on an L-dendriform
algebra $(A,\triangleright, \triangleleft)$, then $\mathcal B$ is
invariant on the L-quadri-algebra given by Eq. (3.34).
\end{prop}

Next we consider the symmetric bilinear forms on an L-quadri-algebra
$(A, \nearrow,  \nwarrow,  \searrow,  \swarrow )$.

\begin{theorem} Let $(A,  \searrow,  \nearrow,  \nwarrow,\
\swarrow)$ be an L-quadri-algebra and $r \in A\otimes A$ be
symmetric. Let $(A,  \triangleright, \triangleleft)$, $(A,  \succ,
\prec)$, $(A,\ \vee, \wedge)$ be the associated horizontal, vertical
and depth L-dendriform algebras of $(A, \searrow,  \nearrow,
\nwarrow,  \swarrow )$ respectively.
 Then the following conditions are equivalent:

(1)\quad $r$ is an $\mathcal {O}$-operator of $(A, \triangleright ,
\triangleleft )$ associated to the bimodule $(L_\ast^*,
-L_\nwarrow^*, -L_\wedge^*, L_\prec^*,  A^*)$.

(2)\quad $r$ is an $\mathcal {O}$-operator of $(A, \succ, \prec)$
associated to the bimodule $(L_\circ^*, R_\nearrow^*,
L_\triangleright^*, -R_\wedge^*,  A^*)$.

(3)\quad $r$ is an $\mathcal {O}$-operator of $(A, \vee, \wedge)$
associated to the bimodule $(L_\bullet^*, R_\swarrow^*,
-L_\triangleleft^*, -R_\prec^*,  A^*)$.

(4)\quad $r$ satisfies
\begin{equation}
r_{13} \triangleright r_{23} = -  (r_{12} \vee r_{23} + r_{12}\wedge
r_{23}) + r_{12} \nwarrow r_{13},
\end{equation}
\begin{equation}
r_{13} \triangleleft r_{23} = (r_{12} \wedge r_{23}) - r_{12} \prec
r_{13}.
\end{equation}

\end{theorem}

\begin{proof} We give the proof of the equivalence between (1) and
(4). The proof of the other equivalences is similar (note that the
symmetry of $r$ is necessary). Let $\{e_1,...,e_n\}$ be a basis of
$A$ and $\{e_1^*,...,e_n^*\}$ be its dual basis. Suppose
$$e_i \searrow e_j = \sum_{k=1} ^n a_{ij}^k e_k,\ e_i \nwarrow e_j =
\sum_{k=1}^n b_{ij}^k e_k,\ e_i \nearrow e_j = \sum_{k=1}^n c_{ij}^k
e_k,\ e_i \swarrow e_j = \sum_{k=1}^n d_{ij}^k e_k,
$$
$$r = \sum_{i,j=1}^na_{ij}e_i \otimes e_j,\ a_{ij} = a_{ji}.
$$
So $r(e_i^*) = \sum_{j=1}^n a_{ij}e_j$. Note that
$$r(a^*) \triangleright r(b^*) = r[L_*^*(r(a^*))b^* - L_\nwarrow^*(r(b^*))a^*],$$ is equivalent to  (for
any $i, j, t$)
$$\sum_{k,l=1}^n a_{ik}a_{jl}(a_{kl}^t - b_{lk}^t) = -\sum_{k,l=1}^n a_{ik}a_{lt}(a_{kl}^j +
c_{kl}^j + b_{kl}^j + d_{kl}^j) + \sum_{k,l=1}^n
a_{jk}a_{lt}b_{kl}^i,$$ which is precisely the coefficient of
$e_i\otimes e_j\otimes e_t$ in
$$r_{13} \triangleright r_{23} = -  (r_{12} \vee r_{23} + r_{12}\wedge
r_{23}) + r_{12} \nwarrow r_{13}.$$ Similarly, we show that
$$r(a^*) \triangleleft r(b^*) = r[-L_\wedge^*(r(a^*))b^* +
L_\prec^*(r(b^*))a^*],$$ is equivalent to
$$
r_{13} \triangleleft r_{23} = (r_{12} \wedge r_{23}) - r_{12} \prec
r_{13}.$$ Hence the conclusion holds.
\end{proof}

\begin{theorem} Let $(A, \searrow,  \nearrow,  \nwarrow,  \swarrow
)$ be an L-quadri-algebra and $r \in A\otimes A$. Suppose that $r$
is symmetric and invertible. Then $r$ satisfies Eqs. (3.39)-(3.40)
 if and only if the nondegenerate bilinear form $\mathcal
{B}$ induced by $r$ through Eq. (1.9) satisfies (for any $x,y,z\in
A$)
\begin{equation}
\mathcal{B}(x\swarrow y, z) = -\mathcal{B}(y, z\bullet x) -
\mathcal{B}(x, z\vee y),\end{equation}\begin{equation}
\mathcal{B}(x\nearrow y, z) = \mathcal{B}(y, x\wedge z - z\vee x)
- \mathcal{B}(x, z\succ y).
\end{equation}
\end{theorem}

\begin{proof}
Let $r=\sum_i a_i\otimes b_i$. Since $r$ is symmetric, we have
$\sum_i a_i\otimes b_i=\sum_ib_i\otimes a_i$. Therefore
$r(v^*)=\sum_i\langle v^*,a_i\rangle b_i=\sum_i\langle
v^*,b_i\rangle a_i$ for any $v^*\in A^*$. Since $r$ is
nondegenerate, for any $x,y,z\in A$, there exist $u^*, v^*, w^*\in
A^*$ such that $x=r(u^*), y=r(v^*), z=r(w^*)$. So

{\small \begin{eqnarray*} {\mathcal B}(x\swarrow y, z)&=& \langle
r(u^*)\swarrow r(v^*), w^*\rangle= \sum_{i,j}\langle u^*,
b_i\rangle\langle v^*, b_j\rangle\langle
w^*, a_i\swarrow a_j\rangle=\langle w^*\otimes u^*\otimes v^*, r_{12}\swarrow r_{13}\rangle;\\
{\mathcal B}(z\bullet x, y)&=& \langle r(w^*)\bullet r(u^*),
v^*\rangle= \sum_{i,j}\langle w^*, b_i\rangle\langle u^*,
b_j\rangle\langle
v^*, a_i\bullet a_j\rangle=\langle w^*\otimes u^*\otimes v^*, r_{13}\bullet r_{23}\rangle;\\
{\mathcal B}(z\vee y, x)&=& \langle r(w^*)\vee r(v^*), u^*\rangle=
\sum_{i,j}\langle w^*, b_i\rangle\langle v^*, b_j\rangle\langle
u^*, a_i\vee a_j\rangle=\langle w^*\otimes u^*\otimes v^*,
r_{12}\vee r_{23}\rangle.
\end{eqnarray*}}
Hence Eq. (3.41) holds if and only if
\begin{equation}r_{13}\bullet r_{23}=-r_{12}\swarrow
r_{13}-r_{12}\vee r_{23},\end{equation} which is exactly the sum
of Eqs. (3.39) and (3.40). Similarly, Eq. (3.40) holds if and only
if
\begin{equation}\mathcal B(x\triangleleft y,z)=\mathcal B(y, x\vee
z)-\mathcal B(x,y\prec z),\;\;\forall x,y,z\in A.\end{equation}
Note that the above equation is exactly the difference of  Eqs.
(3.41) and (3.42): \begin{equation}\mathcal{B}(x\nearrow y, z)-
\mathcal{B}(y\swarrow x, z) =
 \mathcal{B}(y, x\wedge z - z\vee x) - \mathcal{B}(x, z\succ y)+\mathcal{B}(x, z\bullet y)
 +\mathcal{B}(y, z\vee x).\end{equation}
So if Eqs. (3.39) and (3.40) holds, then Eqs. (3.41) and (3.44)
hold. Therefore Eq. (3.42) holds. Conversely, if Eqs. (3.41) and
(3.42) hold, then Eqs. (3.40) and (3.43) hold. Therefore Eq.
(3.39) hold.
\end{proof}

\begin{defn}
{\rm Let $(A,\ \searrow,\ \nearrow, \nwarrow, \swarrow)$ be an
L-quadri-algebra and $r \in A\otimes A$. Eqs. (3.39)-(3.40) are
called $LQ$-equation (a set of equations) in $(A, \searrow,
\nearrow,\nwarrow, \swarrow)$. On the other hand, a symmetric
bilinear form $\mathcal {B}$ on $A$ is called {\it 2-cocycle} if
$\mathcal {B}$ satisfies  Eqs. (3.41)-(3.42). }
\end{defn}

\begin{remark}{\rm
In \cite{NB}, we show that the $LQ$-equation in an
L-quadri-algebra is related to a double construction of the
nondegenerate (skew-symmetric) invariant bilinear forms on
L-quadri-algebras (or equivalently, the double construction of the
nondegenerate (skew-symmetric) 2-cocycles on L-dendriform
algebras). In this sense, the $LQ$-equation in an L-quadri-algebra
can be regarded as an analogue of the CYBE in an L-quadri-algebra.
}\end{remark}

\begin{remark}{\rm Let $(A, \searrow,  \nearrow,  \nwarrow,  \swarrow)$ be an L-quadri-algebra.
Then there exists an L-quadri-algebra structure on $A\oplus {\mathbb
F}c$ given by (for any $x,y\in A$)
\begin{equation}
{\small\begin{array}{rcl} &&x\bar \searrow y= x\searrow y -\mathcal
B(y,x)c,\;\; x\bar \nwarrow y=
x\nwarrow y -\mathcal B(x,y)c, \\
\\
&&x\bar\nearrow y=
x\nearrow y +\mathcal B(y,x)c,\;\; x\bar\swarrow y= x\swarrow y +\mathcal B(x,y)c,\\
\\
&& c\bar\swarrow x=x\bar\swarrow c=c\bar\swarrow c= c\bar \nwarrow
x=x\bar \nwarrow c=c\bar \nwarrow c= x\bar\nearrow c=c\bar\nearrow
x=c\bar\nearrow c=x\bar\swarrow c=c\bar\swarrow x=c\bar\swarrow c=0,
\end{array}}
\end{equation} if and only if
$\mathcal B$ satisfies (for any $x,y,z\in A$)
\begin{equation}
\mathcal {B}(x\nearrow y, z) =  -\mathcal {B}(y, z\vee x-x\wedge
z) - \mathcal {B}(z\succ y,
x),\end{equation}
\begin{equation}\mathcal {B}(x\swarrow y, z) =
-\mathcal {B}(z\triangleright x + z\triangleleft x,y) - \mathcal
{B}(x, z\vee y).
\end{equation}
Furthermore, if a bilinear form $\mathcal B$ on an
L-quadri-algebra $(A, \searrow,  \nearrow,  \nwarrow,  \swarrow)$
satisfying Eqs. (3.47) and (3.48), then it is easy to show that
\begin{equation}
\omega (x,y)=\mathcal B(x,y)-\mathcal B(y,x),\;\;\forall x,y\in A,
\end{equation}
is a 2-cocycle of both the associated vertical L-dendriform
algebra $(A,\succ, \prec)$ and the associated depth L-dendriform
algebra $(A, \vee, \wedge)$. Obviously, if $\mathcal B$ is
symmetric, then Eqs. (3.41)-(3.42) and (3.47)-(3.48)
coincide.}\end{remark}

The following conclusion is obvious.

\begin{prop}
Let $(A, \searrow,  \nearrow,  \nwarrow,  \swarrow )$ be an
L-quadri-algebra. A skew-symmetric (symmetric respectively) bilinear
form $\mathcal {B}$ is invariant (a 2-cocycle respectively) if and
only if $\mathcal {B}$ is invariant (a 2-cocycle respectively)  on
its transpose and symmetries.
\end{prop}

\section*{Appendix: L-octo-algebras}

\setcounter{equation}{0}
\renewcommand{\theequation}
{A.\arabic{equation}}

\noindent {\bf Definition A.1.}\quad  {\rm Let $A$  be a vector
space with eight bilinear products denoted by
$$\searrow_1, \searrow_2, \nearrow_1, \nearrow_2, \nwarrow_1, \nwarrow_2, \swarrow_1,
\swarrow_2:  A\otimes A\rightarrow A.$$ $(A, \searrow_1, \searrow_2,
\nearrow_1, \nearrow_2, \nwarrow_1, \nwarrow_2, \swarrow_1,
\swarrow_2)$ is called an {\it L-octo-algebra} if (for any $x, y, z
\in A$)
\begin{equation}
x\searrow_2(y\searrow_2 z) -(x\ast_{12} y)\searrow_2 z =
y\searrow_2(x\searrow_2 z) -(y\ast_{12}x)\searrow_2 z,
\end{equation}
\begin{equation}
x\searrow_2(y\nearrow_2 z)- (x\vee_{12} y)\nearrow_2 z=y\nearrow_2
(x\succ_2 z)- (y\wedge_{12}x)\nearrow_2 z,
\end{equation}
\begin{equation}
 x\searrow_2 (y\nearrow_1 z)- (x\vee_2y)\nearrow_1 z = y \nearrow_1(x\succ_{12} z)
-(y\wedge_1x)\nearrow_1 z,
\end{equation}
\begin{equation}
x \nearrow_2 (y \succ_1 z)- (x\wedge_2 y) \nearrow_1 z = y
\searrow_1 (x\nearrow_{12} z) -(y\vee_1 x) \nearrow_1 z,
\end{equation}
\begin{equation}
x\searrow_2 (y \searrow_1 z)- (x\ast_2 y) \searrow_1 z = y
\searrow_1 (x\searrow_{12} z)-(y\ast_1 x) \searrow_1 z,
\end{equation}
\begin{equation}
x \searrow_2 (y \nwarrow_1 z)- (x \searrow_2 y ) \nwarrow_1 z =  y
\nwarrow_1 (x \ast_{12} z)- (y \nwarrow_1 x) \nwarrow_1 z,
\end{equation}
\begin{equation}
x\nearrow_2 (y\prec_1 z)- ( x \nearrow_2 y)\nwarrow_1 z   = y
\swarrow_1 (x \wedge_{12} z)-( y \swarrow_1 x)\nwarrow_1 z,
\end{equation}
\begin{equation}
x \nearrow_1 (y\prec_{12} z)-(x \nearrow_1 y) \nwarrow_1 z= y
\swarrow_2 (x\wedge_1 z) - (y \swarrow_2 x) \nwarrow_1 z,
\end{equation}
\begin{equation}
x \searrow_1 (y\swarrow_{12} z)-(x\succ_1 y) \nwarrow_1 z= y
\swarrow_2 (x\vee_1 z) - (y\prec_2 x) \nwarrow_1 z,
\end{equation}
\begin{equation}
x \searrow_1 (y\nwarrow_{12} z)-(x \searrow_1 y ) \nwarrow_1 z =
y\nwarrow_2 (x \ast_1 z)-(y \nwarrow_2 x) \nwarrow_1 z,
\end{equation}
\begin{equation}
x \searrow_2(y \swarrow_1 z)-(x \succ_2 y)\swarrow_1 z  =  y
\swarrow_1 (x \vee_{12} z)- (y \prec_1 x) \swarrow_1 z,
\end{equation}
\begin{equation}
x \searrow_2 (y \swarrow_2 z) - (x\succ_{12}y)\swarrow_2 z=y
\swarrow_2 (x\vee_2 z) -(y\prec_{12} x)\swarrow_2 z,
\end{equation}
\begin{equation}
x\nearrow_2 (y \prec_2 z) - (x\nearrow_{12}y) \nwarrow_2 z=y
\swarrow_2 (x\wedge_2 z) -(y\swarrow_{12}x) \nwarrow_2 z,
\end{equation}
\begin{equation}
x \searrow_2 (y \nwarrow_2 z)-(x\searrow_{12}y) \nwarrow_2 z= y
\nwarrow_2 (x \ast_2 z)- (y\nwarrow_{12}x) \nwarrow_2 z,
\end{equation}
where
\begin{equation}
x\vee_1 y = x \searrow_1 y + x \swarrow_1 y,\ x\wedge_1 y = x
\nearrow_1 y + x \nwarrow_1 y,
\end{equation}
\begin{equation}
x\vee_2 y = x \searrow_2 y + x \swarrow_2 y,\ x\wedge_2 y = x
\nearrow_2 y + x \nwarrow_2 y,
\end{equation}
\begin{equation}
  x\succ_1 y =x \searrow_1 y+ x \nearrow_1 y,  x\prec_1 y
= x \nwarrow_1 y + x \swarrow_1 y,
\end{equation}
\begin{equation}
x\succ_2 y = x \searrow_2 y + x \nearrow_2 y,  x\prec_2 y = x
\nwarrow_2 y+ x \swarrow_2 y,
\end{equation}
\begin{equation}
x\nearrow_{12} y = x \nearrow_1 y + x \nearrow_2 y,\
 x\searrow_{12} y =  x \searrow_1 y + x\searrow_2 y,
\end{equation}
\begin{equation}
x\swarrow_{12} y = x \swarrow_1 y + x \swarrow_2 y,  x\nwarrow_{12}
y = x \nwarrow_1 y + x \nwarrow_2 y.
\end{equation}
\begin{equation}
x \vee_{12} y =x \vee_{1} y+x \vee_{2} y=x\searrow_{12}y
+x\swarrow_{12}y, x\wedge_{12} y = x\wedge_{1} y+x\wedge_{2} y=x
\nearrow_{12} y + x \nwarrow_{12} y,
\end{equation}
\begin{equation}
x \succ_{12} y = x\succ_1 y+x\succ_2 y=x\nearrow_{12} y +
x\searrow_{12} y,\;x\prec_{12} y = x\prec_1 y+x\prec_2
y=x\nwarrow_{12} y + x\swarrow_{12}y,
\end{equation}
\begin{equation}
x\ast_1 y=x \searrow_{1} y + x \nwarrow_{1} y + x \nearrow_{1} y + x
\swarrow_{1} y,\; x\ast_2 y=x \searrow_{2} y + x \nwarrow_{2} y + x
\nearrow_{2} y + x \swarrow_{2} y,
\end{equation}
\begin{equation}
x \ast_{12} y = x\ast_1 y+x\ast_2y= \searrow_{12} y + x
\nwarrow_{12} y + x \nearrow_{12} y + x \swarrow_{12} y.
\end{equation}}

\noindent {\bf Proposition A.2.}\quad  {\it Let $(A, \searrow_1,
\searrow_2,
 \nearrow_1, \nearrow_2, \nwarrow_1, \nwarrow_2, \swarrow_1,
\swarrow_2)$ be an L-octo-algebra.

(1)\quad  $(A,  \vee_2, \wedge_2,\wedge_1, \vee_1)$ is an
L-quadri-algebra.

(2)\quad $(A,  \succ_2, \succ_1,  \prec_1,\prec_2)$ is an
L-quadri-algebra.

(3)\quad $(A,  \searrow_{12}, \nearrow_{12},  \nwarrow_{12},\
\swarrow_{12})$ is an L-quadri-algebra.

(4)\quad If we define (for any $x,y\in A$)
\begin{equation}
 x\triangleright_1^2 y= x \searrow_2 y - y \nwarrow_1 x,\
x\triangleleft_1^2 y = x \nearrow_2 y - y \swarrow_1 x,
\end{equation}
\begin{equation}
x\triangleright_2^1 y = x \searrow_1 y - y \nwarrow_2 x,\
x\triangleleft_2^1 y = x \nearrow_1 y - y \swarrow_2 x,
\end{equation}
then $(A,  \triangleright_1^2, \triangleleft_1^2,\
\triangleleft_2^1,  \triangleright_2^1)$ is an L-quadri-algebra.}

\section*{Acknowledgements}

This work was supported in part by the NSFC (10621101, 10921061), NKBRPC (2006CB805905) and SRFDP
(200800550015).

\end{document}